\documentclass[journal]{IEEEtran}

\usepackage{amsmath,amsfonts}
\usepackage{array}
\usepackage[table]{xcolor}
\usepackage[caption=false,font=footnotesize]{subfig}
\usepackage{textcomp}
\usepackage{placeins}
\usepackage{url}
\usepackage{graphicx}
\usepackage{cite}
\usepackage[hidelinks]{hyperref}

\graphicspath{{figures/}}
\begin{document}

\title{Multi-feed Plane Wave Generator for Compact OTA Multi-Target Emulation With Joint Angle–Delay–Doppler Control}

\author{Le Yu, Guangming Dong, Chunhui Li, and Wei Fan,~\IEEEmembership{Senior Member,~IEEE}%
\thanks{Le Yu, Guangming Dong, and Chunhui Li are with the National Mobile Communications Research Laboratory, School of Information Science and Engineering, Southeast University, Nanjing 210096, China.}%
\thanks{Wei Fan is with the National Mobile Communications Research Laboratory, School of Information Science and Engineering, Southeast University, Nanjing 210096, China, and also with Purple Mountain Laboratories, Nanjing 211111, China.}%
\thanks{Corresponding author: Wei Fan (e-mail: weifan@seu.edu.cn).}}

\markboth{Draft}%
{Yu \MakeLowercase{\textit{et al.}}: PWG-Based Compact OTA Measurement for Multi-Target Sensing}

\maketitle

\begin{abstract}
Over-the-air (OTA) radar target emulation is an important method for evaluating the sensing capabilities of integrated sensing and communication (ISAC) and radar devices. Conventional single-feed compact antenna test ranges (CATRs) are generally limited to a single target direction, while multi-feed angular-synthesis methods commonly used in radar testing still require far-field illumination over the device-under-test (DUT) aperture. To overcome these limitations, this paper proposes a radar target OTA emulation framework based on multi-feed plane wave generator (PWG). The key capability is the simultaneous emulation of spatially close targets from different directions with joint angle--range--Doppler control. Target range is emulated through the corresponding delay, while Doppler and the complex echo coefficient are controlled by a channel emulator. Meanwhile, a multi-input amplitude and phase matrix (APM) maps each independent target-signal path to a dedicated PWG excitation vector that synthesizes the corresponding incident plane wave over the same test zone. This enables multiple target directions to be reproduced as local plane waves at a test distance substantially shorter than the conventional far-field requirement. The framework is experimentally validated at 3~GHz using a real $1\times11$ phased-array DUT. The PWG--DUT separation is 1.6~m, substantially shorter than the 7.2~m Fraunhofer distance of the DUT. At this separation, the synthesized $0^{\circ}$ and $10^{\circ}$ incident waves each exhibit an amplitude peak-to-peak error of 1.0~dB. Across two four-target angle--delay cases and one two-target angle--velocity case, the estimated target parameters show angular deviations within $1^{\circ}$, interpolated delay errors within 2.4~ns, and velocity errors within 0.15~m/s. The results demonstrate compact OTA emulation of spatially close multi-target scenes with joint angle--range--Doppler control for large-aperture sensing DUTs.
\end{abstract}


\section{Introduction}
\IEEEPARstart{R}{adar} target emulation is widely used in the development and testing of radar and integrated sensing and communication (ISAC) systems to evaluate target detection and multi-target discrimination capabilities \cite{liu2022isac,diewald2021vehicles,olutomilayo2021extrinsic}. A controllable target is typically characterized by range, Doppler, a complex echo coefficient, and angle of arrival (AoA). Range, Doppler, and the echo coefficient can be emulated by applying delay, frequency shift, and complex gain to the test signal \cite{diewald2021vehicles,gaudio2020otfs,sobotka2025subclock,yoon2025doppler}. AoA emulation is more challenging because it requires a prescribed spatial wavefront over the device-under-test (DUT) aperture. For a large-aperture phased-array DUT, a nearby radiating source produces wavefront curvature and nonuniform aperture illumination, which can distort the receive-beam response and angle estimation. Therefore, accurate over-the-air (OTA) target emulation requires an incident field that approximates the desired far-field wavefront over the DUT aperture \cite{ieee1492021}.

Conventional target emulation can be implemented conductively, providing convenient and repeatable control of delay, Doppler, and attenuation \cite{wang2025ce_ota_isac}. However, conductive methods bypass the antenna aperture and free-space propagation path and therefore cannot evaluate the actual spatial response of an integrated phased-array DUT \cite{li2025conductive_apm}. Radiated OTA approaches preserve the antenna and propagation path by transmitting the emulated target echoes toward the DUT. A straightforward method varies the incident direction by mechanically repositioning the radiating antenna \cite{asghar2021positioner,buddappagari2021vil}. Although effective, mechanical positioning limits the rate of angular variation and becomes cumbersome when multiple target directions are required.

Electronically controlled multi-antenna and multiprobe techniques have therefore become an important approach to target-angle emulation. Selecting individual radiating antennas provides a set of discrete angular states \cite{scheiblhofer2017lowcost,gadringer2018radar,diewald2022quasi}, while joint excitation of multiple antennas can synthesize virtual target directions between the physical radiator locations \cite{diewald2022aoa,diewald2022two_dimensional}. In particular, coherent excitation of two adjacent antennas enables one-dimensional arbitrary-AoA emulation \cite{diewald2022aoa}, and extension to multiple adjacent channels supports two-dimensional angular synthesis \cite{diewald2022two_dimensional}. In radar testing, triad-array methods further synthesize target directions from three radiating antennas within a prescribed angular region \cite{russell1972rfs,weichao2007rfs}. These techniques substantially improve the flexibility of target-angle emulation. However, they do not remove the plane-wave requirement for a large-aperture DUT. The radiating probes must still be sufficiently distant to provide approximately planar illumination over the DUT aperture. Thus, multiprobe angular synthesis alone does not solve the compact-distance problem for large phased-array DUTs.

Compact antenna test ranges (CATRs) and plane-wave generators (PWGs) provide alternative means of establishing approximately planar fields at distances shorter than the conventional far-field requirement \cite{xiao2024quietzone,tang2022compact_catr}. For target emulation, a conventional single-feed CATR uses a reflector to form a planar-wave quiet zone and provides one incident direction at a time \cite{tang2023angle}. Multiple directions can be obtained by combining several compact-range reflector branches \cite{rowell2020multiple} or by using a multi-feed compact range \cite{ning2020multifeed}. However, such multi-angle CATR configurations increase cost and complexity, while the finite number of reflectors or feeds restricts both the available incident directions.

A PWG instead uses an array of radiating elements with controlled amplitude and phase excitations to synthesize a local plane wave over a prescribed test zone \cite{durso2009pwg}. The theoretical and numerical study in \cite{bucci2013pwg} showed that a PWG can synthesize obliquely incident plane waves and related the maximum incidence angle to the required array size and number of radiating elements. This study, however, was limited to theoretical analysis and numerical synthesis. To obtain larger oblique incidence angles without a large free-space PWG, a hybrid chamber was proposed in which a PWG array is placed inside an overmoded rectangular waveguide and wall reflections are used to produce off-axis plane waves at the DUT \cite{maaskant2021hybrid}. However, PWGs have rarely been applied to target emulation. Their use for simultaneously emulating independent target signals from different directions over the same test zone has not been sufficiently investigated.

To address this measurement gap, this paper proposes a PWG-based compact OTA measurement method for multi-target and multi-angle sensing evaluation. Target delay, Doppler, and echo coefficient are controlled using a channel emulator (CE), whereas target AoAs are reproduced by synthesizing the corresponding incident wavefronts with the PWG. A multi-input amplitude and phase matrix (APM) applies a dedicated PWG excitation vector to each independent target-signal path, allowing different incident directions to be synthesized over the same test zone. The resulting target signals and incident directions are jointly reproduced over the DUT aperture at a test distance shorter than the conventional far-field requirement.

The main contributions of this work are as follows:
\begin{itemize}
\item A compact OTA measurement framework is developed for evaluating multiple targets from different directions. It combines controllable target-signal characteristics with PWG-synthesized incident wavefronts at a distance shorter than the conventional far-field requirement.
\item A measured PWG transfer matrix and a multi-input APM are used to map each independent target-signal path to a calibrated spatial stimulus over a common test zone.
\item The method is experimentally evaluated at 3~GHz using a real $1\times11$ phased-array DUT in a nonanechoic indoor environment. The validation considers both the synthesized-field quality and the deviations of the estimated angle, delay, and velocity from their programmed reference values.
\end{itemize}

The remainder of this paper is organized as follows. Section~II presents target-state mapping and PWG-based target-field synthesis. Section~III describes the experimental characterization, DUT measurement and target-parameter estimation, and measurement results. Section~IV concludes this paper.

\section{PWG-Based Target-Field Synthesis}
\begin{figure*}[!t]
\centering
\includegraphics[width=0.9\textwidth]{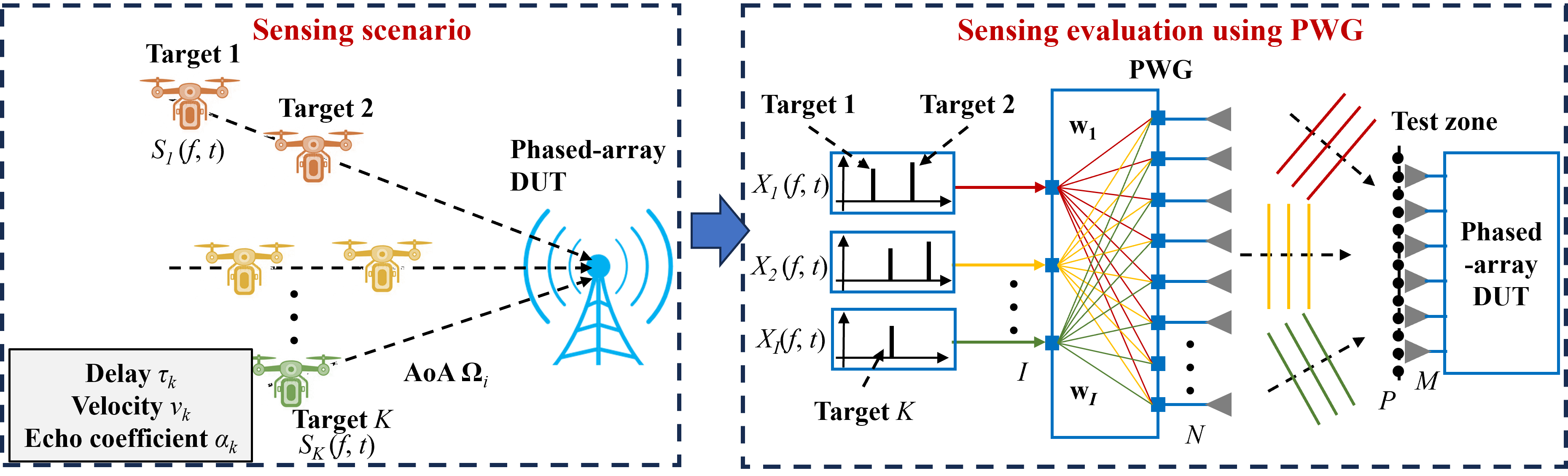}
\caption{Concept and signal flow of the proposed PWG-based OTA target-field synthesis and DUT measurement.}
\label{fig:signal_model}
\end{figure*}

Fig.~\ref{fig:signal_model} illustrates the target-signal mapping and PWG-based spatial-wavefront synthesis. Each emulated target is characterized by a prescribed AoA, round-trip delay, radial velocity, and echo coefficient. Targets sharing the same AoA are combined into an angular-state signal $X_i(f,t)$ and assigned to one independently controlled PWG input path. The corresponding PWG excitation vector synthesizes the desired incident wave over the DUT aperture, and multiple angular-state paths can be activated simultaneously. Section~\ref{subsec:dut_operation} describes the subsequent DUT measurement and target-parameter estimation procedure.

\subsection{Reference Target-State Model and Angular-State Mapping}
As shown in Fig.~\ref{fig:signal_model}, consider $K$ emulated targets whose responses are assigned to the independently controlled PWG input paths. The $k$th target is characterized by its AoA $\boldsymbol{\Omega}_k=(\phi_k,\theta_k)$, round-trip propagation delay $\tau_k$, radial velocity $v_k$, and complex echo coefficient $\alpha_k\in\mathbb C$, where $\phi_k$ and $\theta_k$ denote the azimuth and elevation angles, respectively. The magnitude and phase of $\alpha_k$ specify the target-response magnitude and initial phase, respectively. The corresponding frequency-domain target response at slow time $t$ is
\begin{equation}
S_k(f,t)=\alpha_k e^{-j2\pi f\tau_k}e^{j4\pi f_c v_k t/c},
\label{eq:target_response}
\end{equation}
where $j=\sqrt{-1}$, $f$ is the swept frequency, $f_c$ is the center frequency, and $c$ is the speed of light.
In practical emulation, the CE realizes the Doppler term by varying the phase of its complex channel coefficient according to the prescribed radial velocity.

The $K$ targets are partitioned by AoA. Let $\boldsymbol{\Omega}_i=(\phi_i,\theta_i)$, $i=1,\ldots,I$, denote the $i$th distinct prescribed AoA at the DUT, and let $\mathcal{K}_i$ contain the indices of the targets sharing $\boldsymbol{\Omega}_i$. These targets may have different delays, radial velocities, and echo coefficients. Their responses are combined into the angular-state signal
\begin{equation}
X_i(f,t)=\sum_{k\in\mathcal{K}_i}S_k(f,t),\qquad i=1,\ldots,I.
\label{eq:angular_state_response}
\end{equation}
Each AoA is assigned to one independently controlled PWG input path, so $I$ cannot exceed the number of available input paths. For example, if targets 1 and 2 share $\boldsymbol{\Omega}_1$ while targets 3 and 4 share $\boldsymbol{\Omega}_2$, then $\mathcal K_1=\{1,2\}$, $\mathcal K_2=\{3,4\}$, $X_1=S_1+S_2$, and $X_2=S_3+S_4$.

\subsection{Transfer-Matrix-Based Plane-Wave Synthesis}
As shown in Fig.~\ref{fig:signal_model}, the PWG emulates the AoA of each angular-state signal by synthesizing the corresponding plane wave over the DUT aperture. The angular state $\boldsymbol{\Omega}_i$ specifies the $i$th prescribed AoA at the DUT. Let $\hat{\mathbf{s}}(\boldsymbol{\Omega}_i)$ denote the associated unit propagation vector directed toward the DUT, and let $\mathbf r_p$ denote the position of the $p$th of the $P$ test-zone sampling points. The desired plane-wave vector $\mathbf d_i\in\mathbb C^P$ is defined by
\begin{equation}
\left[\mathbf d_i\right]_p=\exp\{-jk_c\hat{\mathbf{s}}^T(\boldsymbol{\Omega}_i)\mathbf r_p\},\qquad p=1,\ldots,P,
\label{eq:desired_plane_wave}
\end{equation}
where $k_c=2\pi f_c/c$ and the superscript $T$ denotes transpose.

To synthesize $\mathbf d_i$, a PWG with $N$ active elements is used. Let $g_{p,n}\in\mathbb C$ denote the measured co-polar complex transfer coefficient from the $n$th PWG element to the $p$th test-zone sampling point when that element is activated individually. The measured transfer matrix $\mathbf G\in\mathbb C^{P\times N}$ is defined by $[\mathbf G]_{p,n}=g_{p,n}$ for $p=1,\ldots,P$ and $n=1,\ldots,N$. The spatial sampling interval in the test zone is smaller than $\lambda_c/2$, where $\lambda_c=c/f_c$. This sampling density is sufficient to characterize the field distribution. The matrix $\mathbf G$ characterizes the responses from the PWG elements to the test-zone sampling points and includes repeatable effects of the deployed test environment. Because the usable bandwidth in Section~III is small relative to $f_c$, the frequency dependence of this spatial mapping is neglected and the same excitation vectors are applied over that bandwidth.

The continuous PWG excitation vector $\mathbf w_i\in\mathbb C^N$ for the $i$th angular state is obtained from
\begin{equation}
\begin{aligned}
\mathbf{w}_i={}&\underset{\mathbf w}{\operatorname{arg\,min}}\,
\|\mathbf{G}\mathbf{w}-\mathbf{d}_i\|_2^2,\\
&\text{subject to }\|\mathbf w\|_2\leq
\eta\|\mathbf G^{\dagger}\mathbf d_i\|_2,
\end{aligned}
\label{eq:pwg_synthesis}
\end{equation}
where $\mathbf G^{\dagger}$ denotes the Moore--Penrose pseudoinverse of $\mathbf G$, so $\mathbf G^{\dagger}\mathbf d_i$ is the unconstrained minimum-norm least-squares solution. The factor $0<\eta\leq1$ limits the total excitation energy \cite{zhang2023pwg_nonanechoic}. The PWG-side APM excitation vector $\widehat{\mathbf w}_i$ is obtained by scaling $\mathbf w_i$ to the available control range and quantizing it according to the available APM amplitude and phase states. The actual synthesized field vector $\mathbf E_i\in\mathbb C^P$ for the $i$th angular state is
\begin{equation}
\mathbf E_i=\mathbf G\widehat{\mathbf w}_i.
\label{eq:actual_field}
\end{equation}

Each angular state is first synthesized and characterized independently. For the $i$th angular state, let $a_{i,p}=20\log_{10}|[\mathbf E_i]_p|$ and let $\epsilon_{i,p}$ denote the unwrapped phase error relative to the desired plane-wave phase. Let $Q\subseteq\{1,\ldots,P\}$ denote the set of sampling-point indices in the test zone covering the required DUT aperture. The amplitude and phase peak-to-peak errors (PPEs) over $Q$ are
\begin{equation}
\Delta A_i=\max_{p\in Q}a_{i,p}-\min_{p\in Q}a_{i,p},\qquad
\Delta\Phi_i=\max_{p\in Q}\epsilon_{i,p}-\min_{p\in Q}\epsilon_{i,p}.
\label{eq:test_zone_metrics}
\end{equation}
After the field quality of the individual angular states has been confirmed, the independently controlled input paths are activated simultaneously for multi-angle target emulation. Assuming linear and coherent superposition among the independently controlled PWG paths, the total target-field vector over the test-zone samples is
\begin{equation}
\mathbf E_{\mathrm{TZ}}(f,t)=\sum_{i=1}^{I}X_i(f,t)\mathbf E_i
\approx\sum_{i=1}^{I}X_i(f,t)\mathbf d_i.
\label{eq:total_tz_field}
\end{equation}
Equation~\eqref{eq:total_tz_field} shows that each input signal retains its programmed delay, velocity, and echo coefficient while being mapped to the corresponding PWG-synthesized wavefront.

\section{Experimental Characterization and Measurement}
\subsection{Experimental System and Linearity Characterization}
\begin{figure}[!t]
\centering
\includegraphics[width=0.95\columnwidth]{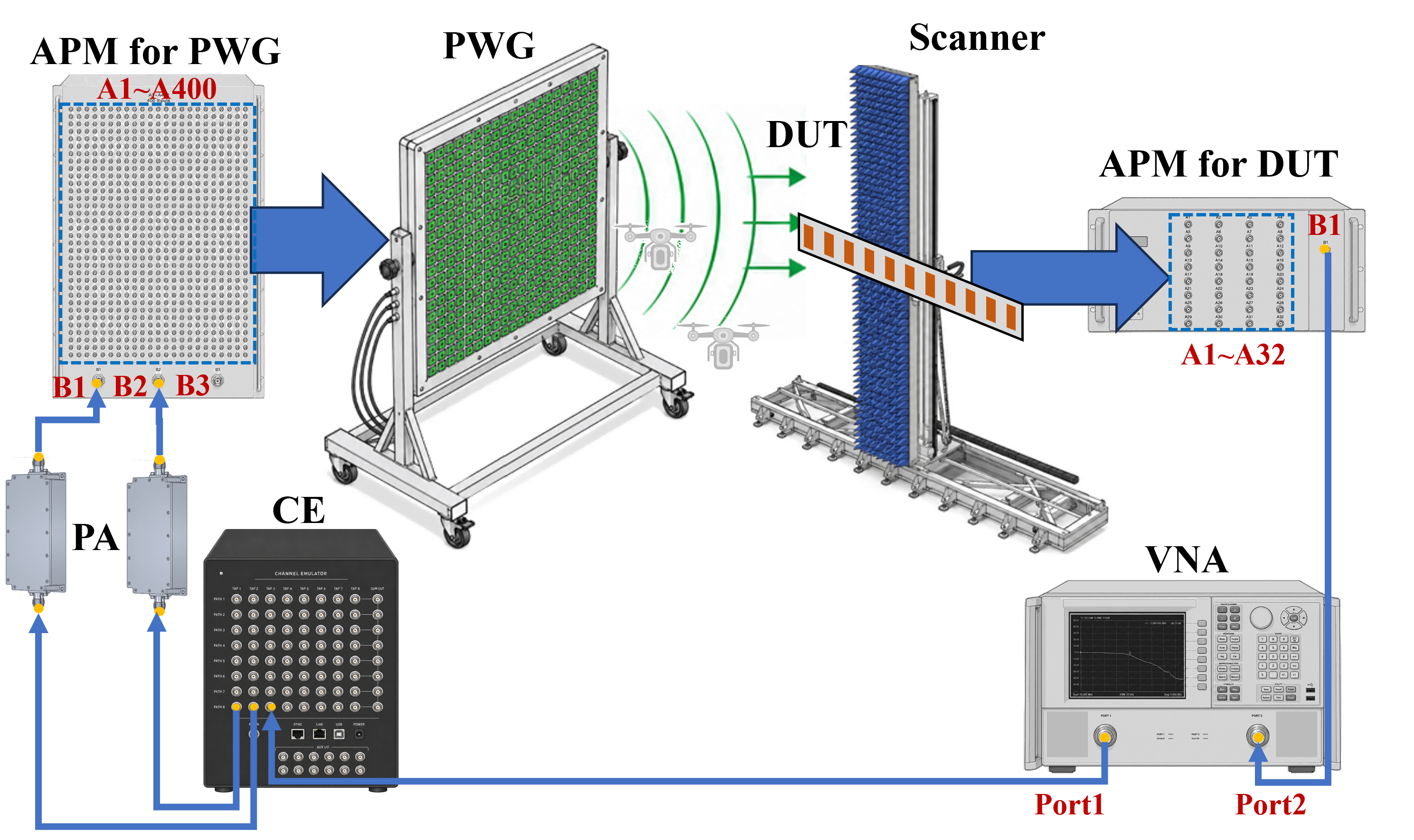}
\caption{Architecture of the PWG-based OTA measurement system.}
\label{fig:system}
\end{figure}

Fig.~\ref{fig:system} shows the hardware implementation of the proposed PWG-based OTA measurement system. The system realizes the complete measurement chain from target-signal generation and PWG-based spatial-wavefront synthesis to DUT beamforming and response acquisition. Port~1 of a vector network analyzer (VNA) supplies the test signal to the CE. The two CE outputs are connected through separate power amplifiers (PAs) to inputs B1 and B2 of the PWG-side APM, so the two angular-state signals can be controlled independently. Each active APM output is connected to one selected PWG element, and this APM-output--PWG-element pair is referred to as one PWG excitation path. On the receive side, the DUT element outputs are routed to the DUT-side APM. The APM loads the calibrated receive weights for each commanded beam state, and VNA port~2 records the resulting combined complex response.

\begin{figure}[!t]
\centering
\includegraphics[width=0.95\columnwidth]{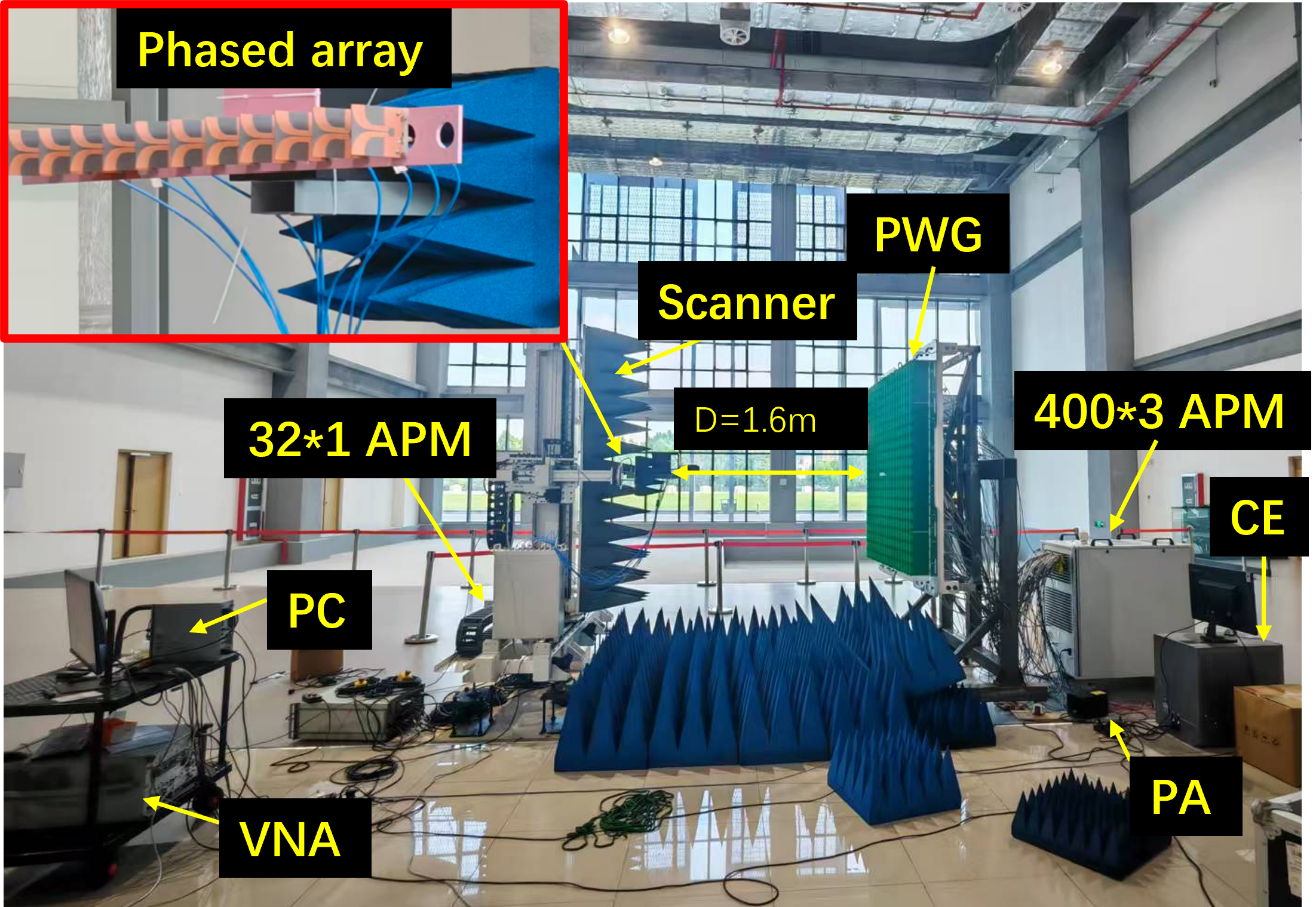}
\caption{Open-laboratory measurement scenario. The inset shows the real $1\times11$ phased-array DUT mounted in front of the local absorber treatment.}
\label{fig:scenario}
\end{figure}

Table~\ref{tab:parameters} summarizes the instrument settings and test parameters. In the transmit section, 100 PWG-side APM output channels drive an active $10\times10$ subarray formed by selecting every other element of the physical $20\times20$ PWG, giving a 15~cm effective element spacing. In the receive section, the synthesized fields illuminate a real $1\times11$ phased-array DUT with a 0.6~m aperture, and the DUT-side APM weights and combines its 11 element outputs. The VNA performs frequency-swept acquisition for the angle--delay measurements and repeated single-frequency acquisition for the angle--velocity measurements. Detailed frequency sampling and intermediate-frequency (IF) bandwidth settings are given in Table~\ref{tab:parameters}. At 3~GHz, the PWG--DUT separation is 1.6~m, which is substantially shorter than the 7.2~m Fraunhofer distance of the DUT. The DUT is therefore evaluated under a calibrated local plane-wave condition produced by the PWG. The open-laboratory arrangement and local absorber treatment are shown in Fig.~\ref{fig:scenario}. A control computer programs both APMs, the CE, and the VNA. The measurement proceeds from multichannel linearity characterization to PWG transfer-matrix identification and field evaluation, DUT receive-path calibration, and finally angle--delay and angle--velocity target measurements.

\begin{table}[!t]
\caption{Instrument Settings and Test Parameters}
\label{tab:parameters}
\centering
\footnotesize
\renewcommand{\arraystretch}{1.18}
\begin{tabular}{>{\raggedright\arraybackslash}p{0.48\columnwidth}|>{\raggedright\arraybackslash}p{0.42\columnwidth}}
\hline\hline
\textbf{Parameter} & \textbf{Value} \\
\hline
\multicolumn{2}{c}{\textit{Vector Network Analyzer}} \\ \hline
Angle--delay acquisition & 2.8--3.2~GHz, 1001 points, 100~Hz IF bandwidth \\ \hline
Angle--velocity acquisition & Complex $S_{21}$ at 3~GHz, 10~kHz IF bandwidth \\ \hline
Equivalent slow-time sampling & 2~kHz grid, 2001 samples over 1~s \\ \hline
\multicolumn{2}{c}{\textit{Channel Emulator}} \\ \hline
Active port configuration & 1 input and 2 outputs \\ \hline
Bandwidth & 80~MHz \\ \hline
\multicolumn{2}{c}{\textit{PWG-Side APM}} \\ \hline
Active configuration & 2 of 3 inputs and 100 of 400 outputs \\ \hline
Phase control & 8 bit ($\sim1.4^{\circ}$ step) and $\pm4^{\circ}$ typ. accuracy \\ \hline
Attenuation control & $-40$ to 0~dB operating range, $-90$~dB off state, 0.5~dB step, and $\pm1$~dB typ. accuracy \\ \hline
\multicolumn{2}{c}{\textit{Plane Wave Generator}} \\ \hline
Physical array & $20\times20$ Vivaldi elements with 7.5~cm spacing \\ \hline
Operating band and polarization & 1--4~GHz, vertical polarization \\ \hline
Active subarray & $10\times10$ elements selected at 15~cm effective spacing \\ \hline
Active aperture & 1.35~m $\times$ 1.35~m \\ \hline
\multicolumn{2}{c}{\textit{DUT Receive Chain}} \\ \hline
DUT array / aperture & $1\times11$ / 0.6~m \\ \hline
Fraunhofer distance & Approximately 7.2~m at 3~GHz \\ \hline
DUT-side APM & 8-bit phase ($\sim1.4^{\circ}$) and 0.25~dB amplitude step \\ \hline
\multicolumn{2}{c}{\textit{Test Arrangement}} \\ \hline
Separation / field scan & 1.6~m / $41\times41$ points at 30~mm spacing \\ \hline
Environment & Open laboratory with local absorber \\
\hline\hline
\end{tabular}
\end{table}

To characterize the linear and coherent superposition across the PWG excitation paths required by \eqref{eq:total_tz_field} and identify potential interference in the open-laboratory environment, a preliminary frequency-sweep test was conducted. With the probe fixed at the test-zone center, the 100 APM--PWG excitation paths were sequentially switched on and off over 2--4~GHz to acquire their individual transfer responses. A preset weight vector was then applied to these responses in postprocessing to predict the simultaneous response. Finally, the same quantized weights were loaded into the APM, and all 100 excitation paths were activated simultaneously to measure the actual response.

\begin{figure}[!b]
\centering
\includegraphics[width=0.9 \columnwidth]{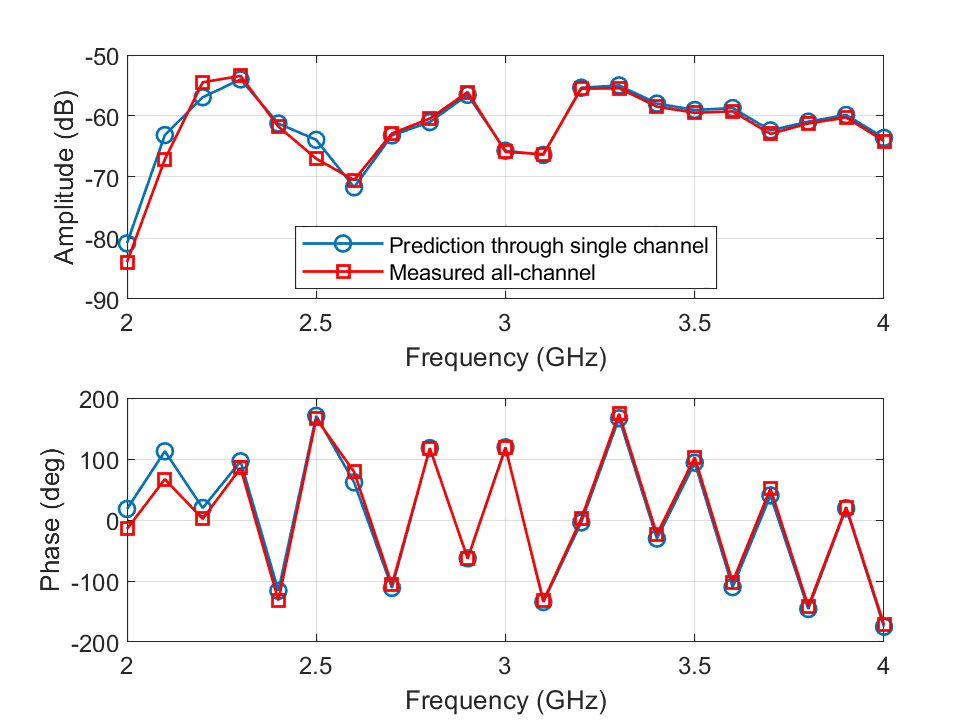}
\caption{Predicted response obtained from individual excitation-path measurements and the corresponding simultaneous 100-path measurement at the test-zone center.}
\label{fig:frequency_sweep}
\end{figure}

Fig.~\ref{fig:frequency_sweep} compares the postprocessed prediction with the simultaneous 100-path measurement. This comparison directly evaluates whether the response predicted from the individually measured excitation paths is retained when all paths operate together. The measurement shows appreciable discrepancies below approximately 2.6~GHz, suggesting that the open laboratory contains additional interference in this frequency range. In contrast, the predicted and measured amplitude and phase responses nearly coincide from 2.8 to 3.2~GHz. The close agreement characterizes stable linear and coherent synthesis across the excitation paths over this frequency range and suggests that residual APM interport coupling is sufficiently small in the present configuration. Accordingly, 3~GHz was selected for the subsequent spatial-field and target measurements.

\subsection{Spatial-Field Characterization and DUT Calibration}
\begin{figure*}[!t]
\centering
\subfloat[]{\includegraphics[width=0.32\textwidth]{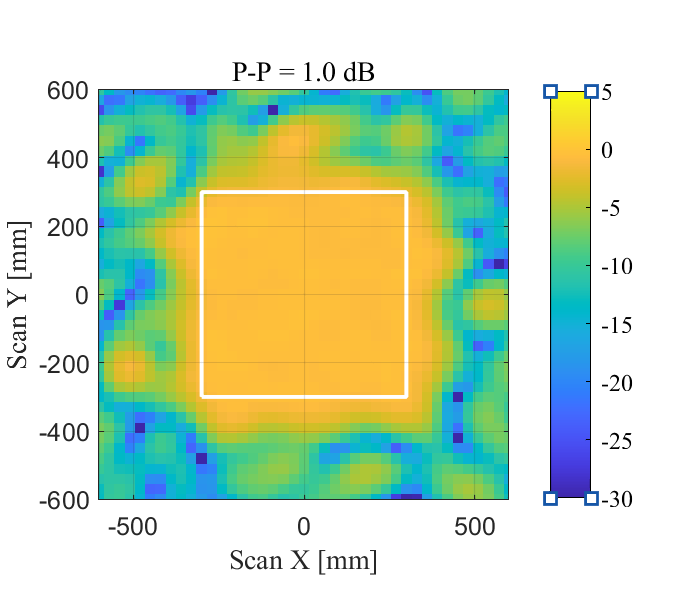}}
\hfill
\subfloat[]{\includegraphics[width=0.32\textwidth]{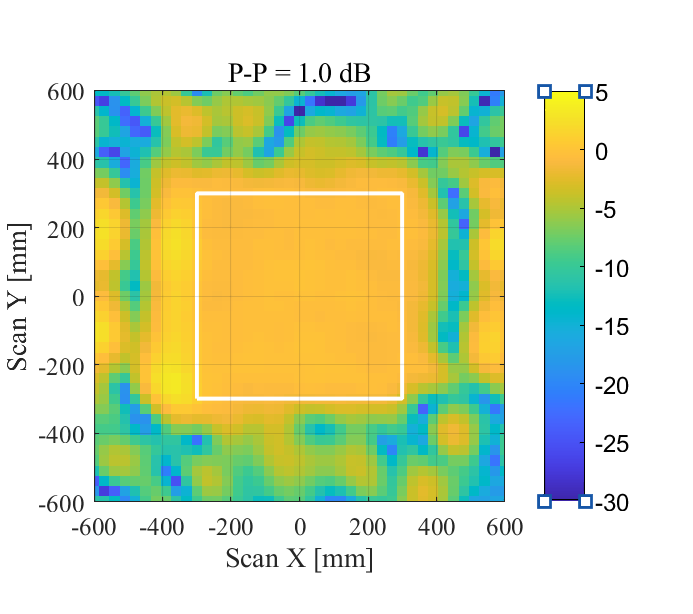}}
\hfill
\subfloat[]{\includegraphics[width=0.32\textwidth]{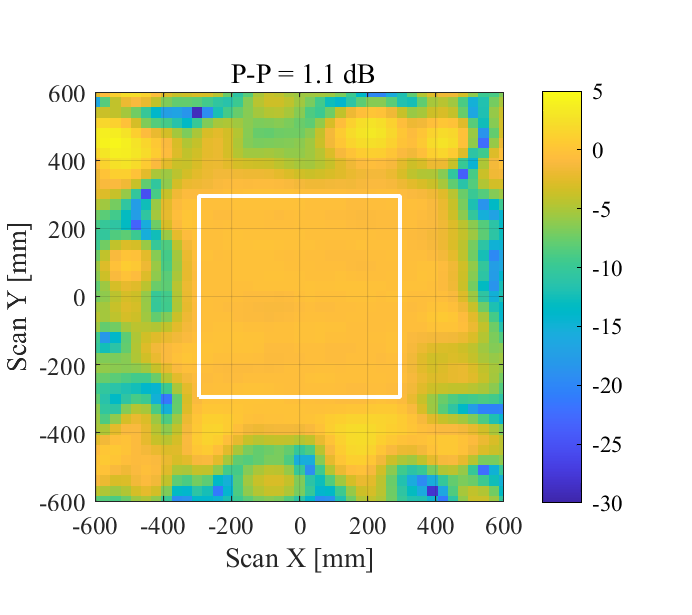}}\\[-0.5ex]
\subfloat[]{\includegraphics[width=0.32\textwidth]{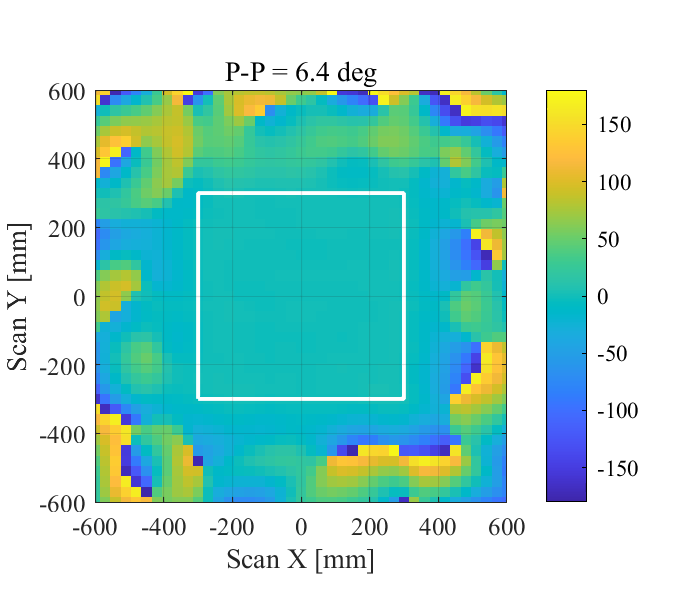}}
\hfill
\subfloat[]{\includegraphics[width=0.32\textwidth]{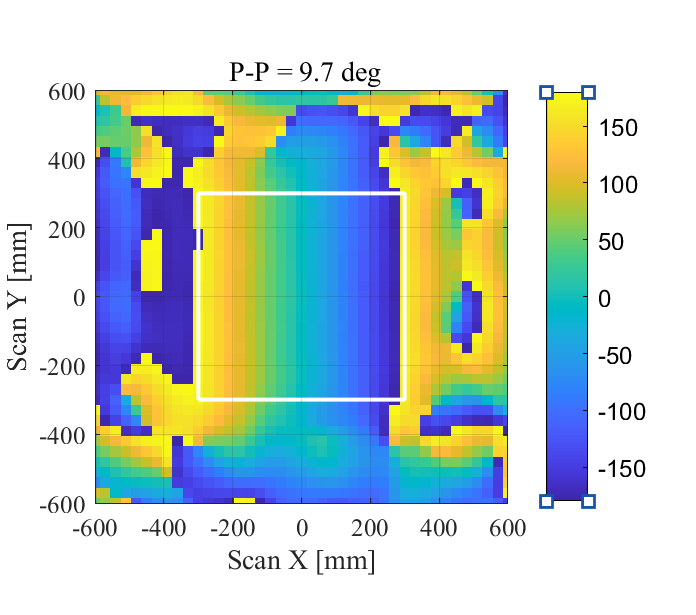}}
\hfill
\subfloat[]{\includegraphics[width=0.32\textwidth]{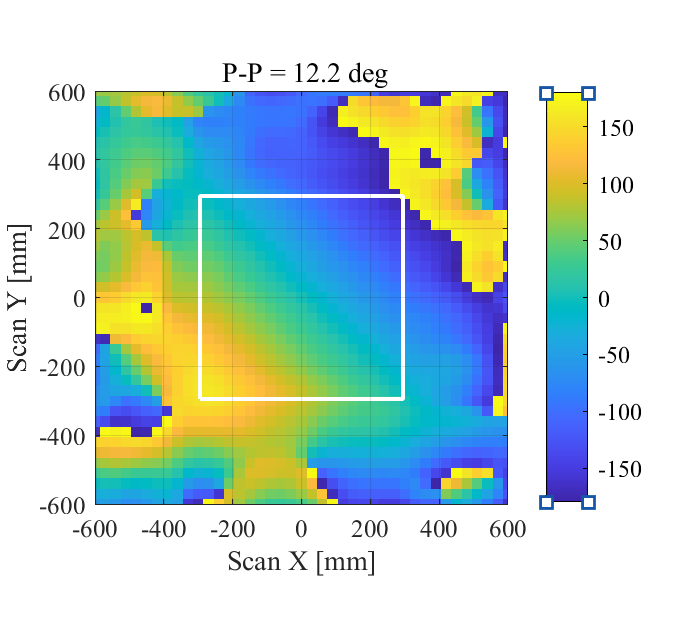}}
\caption{Spatial-field characterization of the PWG at 3~GHz: (a) amplitude and (d) phase for $(\phi,\theta)=(0^{\circ},0^{\circ})$, (b) amplitude and (e) phase for $(10^{\circ},0^{\circ})$, and (c) amplitude and (f) phase for $(5^{\circ},5^{\circ})$. The white rectangles delineate the test zone used for PPE evaluation.}
\label{fig:pwg_fields}
\end{figure*}

The purpose of this subsection is to identify the transfer matrix $\mathbf{G}$ at the 3~GHz center frequency, as required by Section~II-B, and characterize the corresponding PWG-synthesized fields. The measured signal-to-noise ratio is approximately 40~dB. The $-90$~dB attenuation state of the PWG-side APM therefore suppresses an output path below the measurement noise and is used as the off state. A co-polar probe with fixed configuration scans a $1.2\times1.2$~m plane on a 30~mm grid, giving $41\times41$ sampling positions. At each position, one of the 100 active PWG excitation paths is enabled while the remaining paths are set to $-90$~dB. Consequently, constructing $\mathbf{G}$ requires $41\times41\times100=168{,}100$ complex-response measurements and approximately 16~h. The measured $\mathbf{G}$ is then used in \eqref{eq:pwg_synthesis} to obtain the continuous PWG excitation vector $\mathbf w_i$ for each desired AoA. After scaling and quantization, the implemented PWG-side APM excitation vector $\widehat{\mathbf w}_i$ is loaded. Each angular state is synthesized and characterized independently by activating all 100 PWG excitation paths simultaneously. The probe then repeats the $41\times41$ scan to directly measure the synthesized field shown in Fig.~\ref{fig:pwg_fields}. Each characterization scan requires approximately 1~h.

Within the test zone outlined by the white rectangle in each panel of Fig.~\ref{fig:pwg_fields}, the amplitude PPEs defined in \eqref{eq:test_zone_metrics} are 1.0, 1.0, and 1.1~dB for $(\phi,\theta)=(0^{\circ},0^{\circ})$, $(10^{\circ},0^{\circ})$, and $(5^{\circ},5^{\circ})$, respectively. The corresponding phase PPEs are 6.4$^{\circ}$, 9.7$^{\circ}$, and 12.2$^{\circ}$. These spatial-field measurements show that the measured-transfer-matrix optimization produces the required normal-incidence and tilted plane waves in the open laboratory. The residual field distortion outside the marked test zone is attributed to the finite PWG aperture, APM quantization, and environmental reflections.

\begin{figure}[!t]
\centering
\includegraphics[width=0.9\columnwidth]{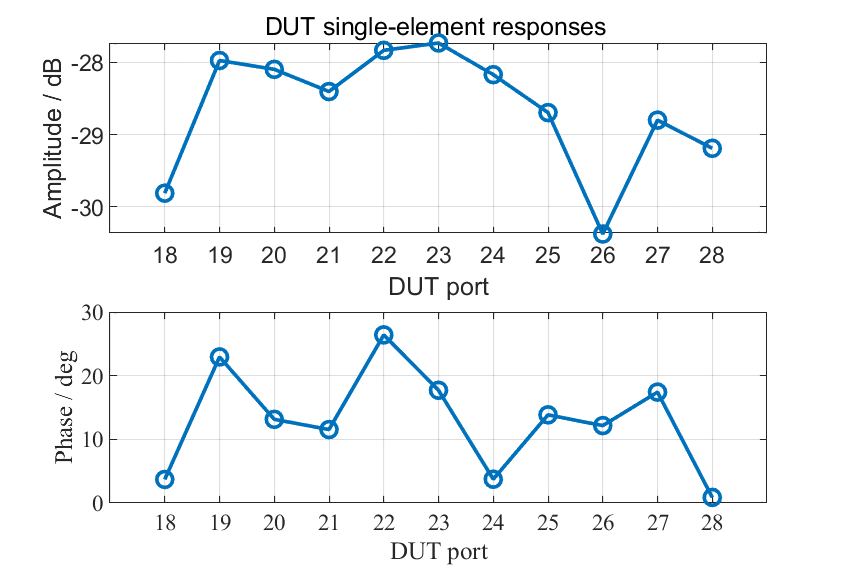}
\caption{Measured complex responses of the 11 DUT elements at 3~GHz used to calibrate the element receive paths.}
\label{fig:dut_calibration}
\end{figure}

Before the target measurements, the 11 DUT element receive paths are calibrated using the synthesized normal-incidence plane wave. The corresponding PWG excitation vector is applied through input B1, and the DUT-side APM sequentially selects the 11 array elements while the VNA records their complex responses. The element with the lowest received power is selected as the reference. This choice allows the remaining higher-power paths to be equalized by attenuation alone within the DUT-side APM. Relative amplitude and phase calibration coefficients are obtained from the measured complex responses. Fig.~\ref{fig:dut_calibration} shows the measured element responses. The resulting calibration coefficients are applied throughout the subsequent angle--delay and angle--velocity measurements to compensate amplitude and phase imbalance during electronic receive-beam steering.

\subsection{DUT Measurement and Target-Parameter Estimation}
\label{subsec:dut_operation}
Following PWG field synthesis and DUT receive-path calibration, the target signals generated by the CE are applied to their assigned PWG input paths. For each target configuration, the calibrated receive weights are successively loaded into the DUT-side APM to steer the DUT receive beam over the commanded azimuth angles. The VNA records the combined complex response at each beam state, and the resulting data are processed to estimate the target parameters. The programmed AoA, delay, and velocity are treated as reference values throughout this paper. The deviations reported in Tables~\ref{tab:angle_delay_results} and \ref{tab:angle_velocity_results} are calculated relative to these programmed values and do not represent absolute measurement accuracy or measurement uncertainty.

Let $\phi_b$ denote the receive-beam steering angle. The receive-beamformer output is
\begin{equation}
\begin{aligned}
Y(\phi_b,f,t)
={}&\sum_{m=1}^{M}u_m(\phi_b)
\sum_{i=1}^{I}h_{i,m}(f)X_i(f,t)+\xi(\phi_b,f,t)\\
={}&\mathbf u^T(\phi_b)\sum_{i=1}^{I}\mathbf h_i(f)X_i(f,t)
+\xi(\phi_b,f,t),
\end{aligned}
\label{eq:received_response}
\end{equation}
where $h_{i,m}(f)$ is the complex response at the $m$th DUT element port produced by a unit angular-state signal incident from $\boldsymbol{\Omega}_i$, and $\mathbf h_i(f)=[h_{i,1}(f),\ldots,h_{i,M}(f)]^T\in\mathbb C^M$ is the corresponding DUT element-response vector. The coefficient $u_m(\phi_b)$ is the complex receive weight directly applied to the $m$th DUT element signal, and $\mathbf u(\phi_b)=[u_1(\phi_b),\ldots,u_M(\phi_b)]^T\in\mathbb C^M$ is the receive-beamforming weight vector. The term $\xi$ denotes the residual noise. Define the beamformed DUT response to the $i$th angular state as $H_{\mathrm{DUT}}(\phi_b,\boldsymbol{\Omega}_i,f)=\mathbf u^T(\phi_b)\mathbf h_i(f)$. The receive-beamformer output can then be written as
\begin{equation}
Y(\phi_b,f,t)=\sum_{i=1}^{I}H_{\mathrm{DUT}}(\phi_b,\boldsymbol{\Omega}_i,f)X_i(f,t)+\xi(\phi_b,f,t).
\label{eq:beam_response_form}
\end{equation}
Equation~\eqref{eq:beam_response_form} links each emulated target signal to the incident AoA generated by the PWG and the beamformed response of the real DUT. The calibrated and quantized receive weights are loaded successively so that the response is sampled over $\phi_b$ for the subsequent transforms.

For static targets, $v_k=0$ and the slow-time dependence vanishes. For angle--delay processing, a frequency window $W_f(f)$ defined over the usable CE bandwidth is applied before the response is transformed into the delay domain,
\begin{equation}
y(\phi_b,\tau)=\mathcal{F}^{-1}_{f}\{W_f(f)Y(\phi_b,f)\}.
\label{eq:angle_delay_transform}
\end{equation}
For dynamic targets, the slow-time response at the center frequency is transformed to obtain the angle--Doppler response,
\begin{equation}
z(\phi_b,f_D)=\mathcal{F}_{t}\{Y(\phi_b,f_c,t)\},\qquad
v=\frac{cf_D}{2f_c}.
\label{eq:angle_velocity_transform}
\end{equation}
Here, $f_D$ is the Doppler-frequency variable, $\mathcal{F}^{-1}_{f}$ denotes the inverse Fourier transform over swept frequency, and $\mathcal{F}_{t}$ denotes the Fourier transform over slow time. The angle--delay and angle--velocity maps are obtained from the normalized powers $|y(\phi_b,\tau)|^2$ and $|z(\phi_b,f_D)|^2$, respectively.

\subsection{Angle--Delay Target Measurement}
For the angle--delay experiment, the CE limits the usable system bandwidth to $B=80$ MHz, corresponding to a nominal delay resolution of $\Delta\tau=1/B=12.5$~ns. The test system emulates two cases, each containing four equal-nominal-power targets. In Case I, T1--T4 are set to $(\phi,\tau)=(0^{\circ},100~\mathrm{ns})$, $(0^{\circ},200~\mathrm{ns})$, $(10^{\circ},100~\mathrm{ns})$, and $(10^{\circ},300~\mathrm{ns})$, respectively. This case includes targets that share either an angle or a delay and therefore tests separability in a partially coupled angle--delay scene. In Case II, the programmed states are $(0^{\circ},100~\mathrm{ns})$, $(0^{\circ},300~\mathrm{ns})$, $(10^{\circ},200~\mathrm{ns})$, and $(10^{\circ},400~\mathrm{ns})$, providing greater separation in both dimensions. The $0^{\circ}$ and $10^{\circ}$ echoes are assigned to inputs B1 and B2 of the PWG-side APM, respectively. At the receiver, the calibrated DUT receive beam is electronically steered from $-40^{\circ}$ to $40^{\circ}$ in $1^{\circ}$ steps. At each beam state, the VNA records the complex frequency response. After frequency windowing, an inverse Fourier transform over the usable 80~MHz band produces the receive-beam angle--delay response according to \eqref{eq:angle_delay_transform}. Each two-dimensional local maximum provides the target-angle estimate through its commanded DUT beam state, while the target delay is obtained from the interpolated delay-domain peak.

\begin{figure}[!t]
\centering
\subfloat[]{\includegraphics[width=0.8\columnwidth]{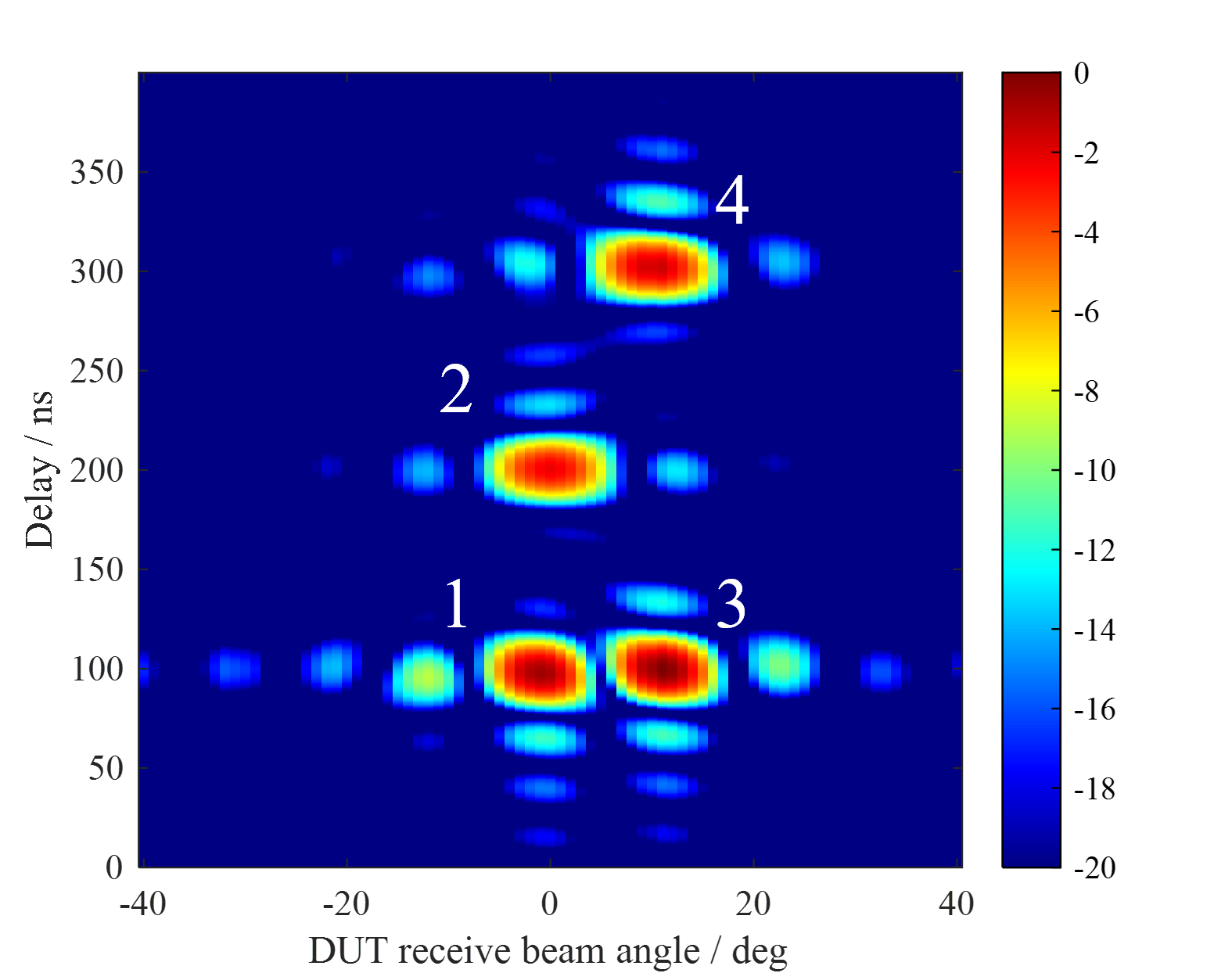}}\\[-0.5ex]
\subfloat[]{\includegraphics[width=0.8\columnwidth]{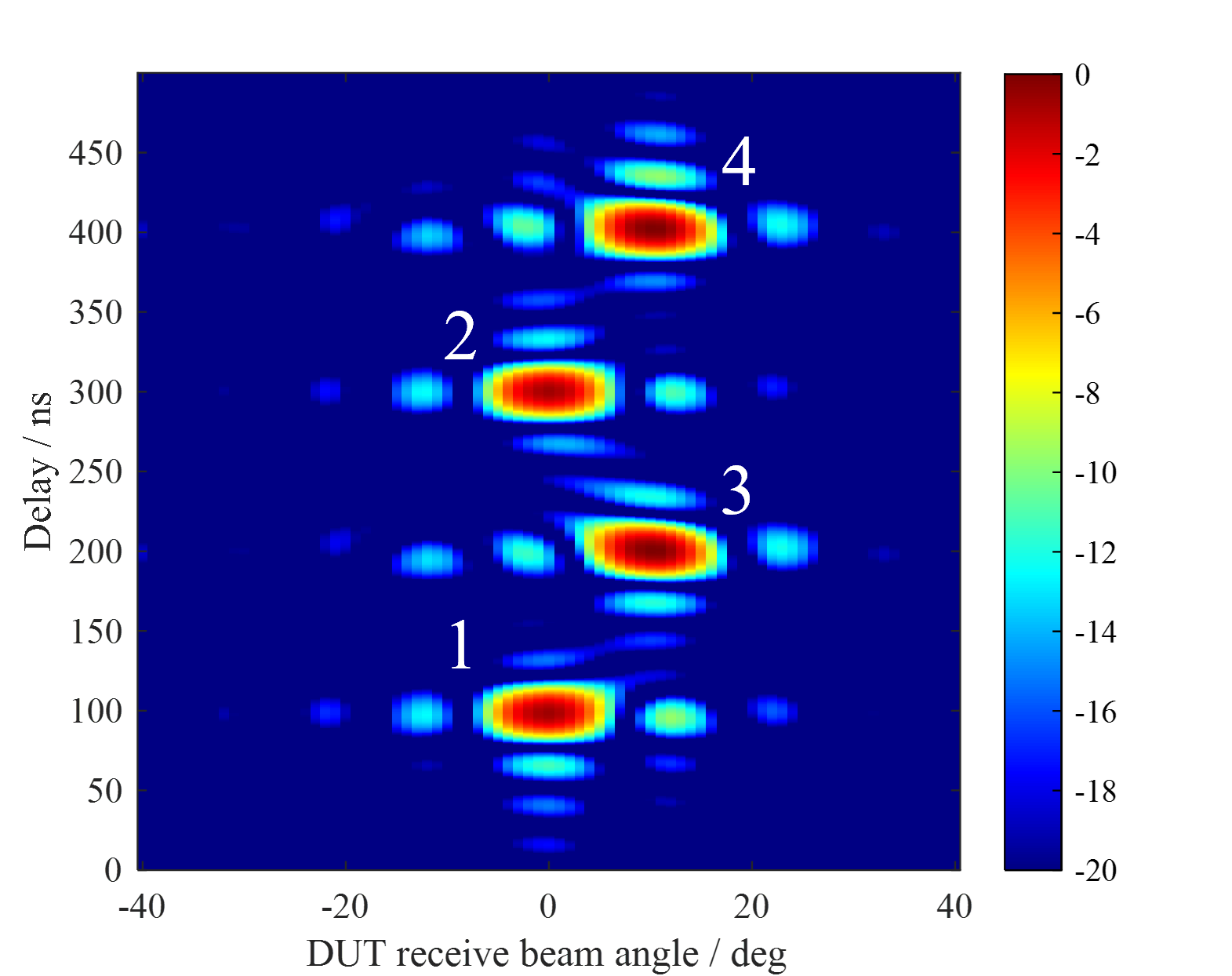}}
\caption{Measured DUT receive-beam angle--delay maps for two emulated target cases: (a) Case I and (b) Case II. Each case contains four targets, and labels 1--4 correspond to targets T1--T4.}
\label{fig:angle_delay_maps}
\end{figure}

\begin{table}[!t]
\centering
\caption{Programmed and Estimated Angle--Delay Target Parameters}
\label{tab:angle_delay_results}
\footnotesize
\setlength{\tabcolsep}{3pt}
\renewcommand{\arraystretch}{1.22}
\resizebox{\columnwidth}{!}{%
\begin{tabular}{c|c|c|c|c}
\hline\hline
\textbf{Target} & \shortstack{\textbf{Programmed}\\$(\phi,\tau)$} & \shortstack{\textbf{Estimated}\\$(\hat{\phi},\hat{\tau})$} & \shortstack{\textbf{Deviation}\\$(\Delta\phi,\Delta\tau)$} & \shortstack{\textbf{Rel. Peak}\\\textbf{Power}} \\
\hline
\multicolumn{5}{c}{\textit{Case I}} \\
\hline
T1 & $(0^{\circ},100~\mathrm{ns})$ & $(-1^{\circ},97.9~\mathrm{ns})$ & $(-1^{\circ},-2.1~\mathrm{ns})$ & $-0.68$~dB \\ \hline
T2 & $(0^{\circ},200~\mathrm{ns})$ & $(0^{\circ},200.6~\mathrm{ns})$ & $(0^{\circ},+0.6~\mathrm{ns})$ & $-2.19$~dB \\ \hline
T3 & $(10^{\circ},100~\mathrm{ns})$ & $(11^{\circ},99.8~\mathrm{ns})$ & $(+1^{\circ},-0.2~\mathrm{ns})$ & $0.00$~dB \\ \hline
T4 & $(10^{\circ},300~\mathrm{ns})$ & $(11^{\circ},302.0~\mathrm{ns})$ & $(+1^{\circ},+2.0~\mathrm{ns})$ & $-0.67$~dB \\
\hline
\multicolumn{5}{c}{\textit{Case II}} \\
\hline
T1 & $(0^{\circ},100~\mathrm{ns})$ & $(0^{\circ},98.5~\mathrm{ns})$ & $(0^{\circ},-1.5~\mathrm{ns})$ & $-0.03$~dB \\ \hline
T2 & $(0^{\circ},300~\mathrm{ns})$ & $(0^{\circ},300.6~\mathrm{ns})$ & $(0^{\circ},+0.6~\mathrm{ns})$ & $-0.60$~dB \\ \hline
T3 & $(10^{\circ},200~\mathrm{ns})$ & $(10^{\circ},200.7~\mathrm{ns})$ & $(0^{\circ},+0.7~\mathrm{ns})$ & $0.00$~dB \\ \hline
T4 & $(10^{\circ},400~\mathrm{ns})$ & $(10^{\circ},402.4~\mathrm{ns})$ & $(0^{\circ},+2.4~\mathrm{ns})$ & $-0.03$~dB \\
\hline\hline
\end{tabular}}
\end{table}

The measured angle--delay maps for Cases I and II are shown in Fig.~\ref{fig:angle_delay_maps}(a) and Fig.~\ref{fig:angle_delay_maps}(b), respectively. The relative peak powers reported in Table~\ref{tab:angle_delay_results} are normalized to the strongest estimated target in the corresponding case. Table~\ref{tab:angle_delay_results} shows that Case I has delay deviations no larger than 2.1~ns and angular deviations of at most $1^{\circ}$. The observed peak shifts are mainly associated with the finite angular resolution of the real $1\times11$ DUT and the interaction between closely spaced target responses. For the same reason, equal nominal target powers do not necessarily produce equal estimated peak powers, as illustrated by the $-2.19$~dB value for T2. The relative peak powers therefore characterize the measured DUT response and should not be interpreted as the echo-coefficient accuracy of the CE. Residual PWG synthesis and measurement errors may also contribute. In Case II, all estimated angles match the programmed states, and the delay deviations remain within 2.4~ns. This result demonstrates the expected PWG-to-DUT angular mapping when the target peaks are well separated. The residual delay deviations are small relative to the 12.5~ns nominal resolution and reflect peak interpolation and processing-window effects.

\subsection{Angle--Velocity Target Measurement}
For the angle--velocity experiment, two equal-nominal-power targets are programmed at $(\phi,v)=(0^{\circ},5~\mathrm{m/s})$ and $(10^{\circ},10~\mathrm{m/s})$, corresponding to Doppler frequencies of 100 and 200~Hz at 3~GHz, respectively. The two target signals are assigned to inputs B1 and B2 of the PWG-side APM. The CE generates the prescribed Doppler signatures through time-varying complex channel coefficients. The calibrated DUT electronically scans its receive beam from $-40^{\circ}$ to $40^{\circ}$ in $2^{\circ}$ steps. At each beam angle, the CE playback and VNA acquisition are synchronized to emulate a 2~kHz slow-time sampling grid over 1~s, yielding 2001 complex samples. The 2~kHz value denotes the equivalent sampling rate rather than the physical acquisition rate of the VNA. The measured slow-time sequences are Fourier transformed according to \eqref{eq:angle_velocity_transform} to obtain the angle--velocity response. The commanded beam angle of each local maximum gives the target-angle estimate, and its Doppler peak is converted to radial velocity using \eqref{eq:angle_velocity_transform}.

\begin{figure}[!t]
\centering
\includegraphics[width=0.9\columnwidth]{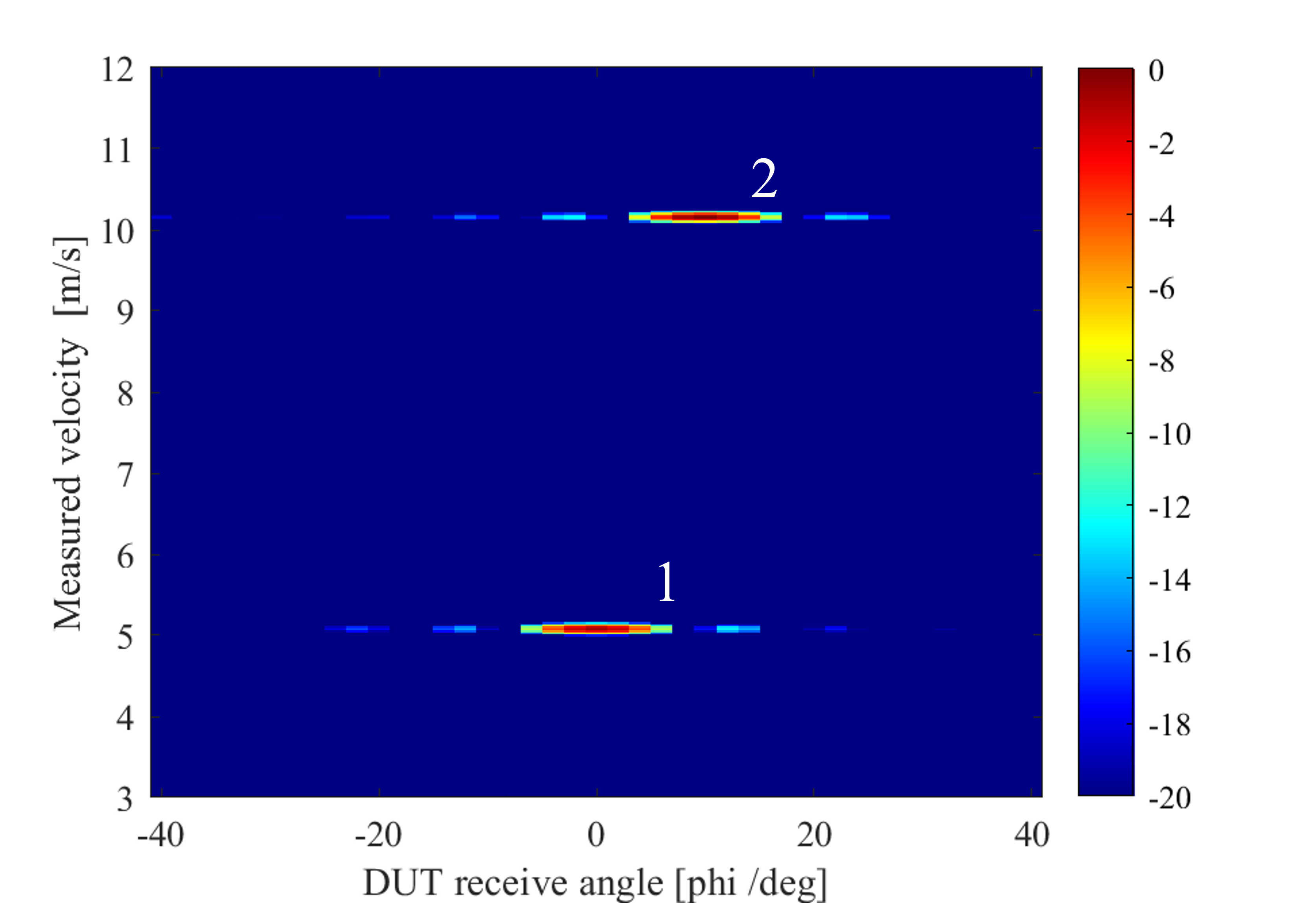}
\caption{Measured DUT receive-beam angle--velocity map for two emulated moving targets programmed at $(0^{\circ},5~\mathrm{m/s})$ and $(10^{\circ},10~\mathrm{m/s})$.}
\label{fig:angle_velocity}
\end{figure}

\begin{table}[!t]
\centering
\caption{Programmed and Estimated Angle--Velocity Target Parameters}
\label{tab:angle_velocity_results}
\footnotesize
\setlength{\tabcolsep}{3pt}
\renewcommand{\arraystretch}{1.22}
\resizebox{\columnwidth}{!}{%
\begin{tabular}{c|c|c|c|c}
\hline\hline
\textbf{Target} & \shortstack{\textbf{Programmed}\\$(\phi,v)$} & \shortstack{\textbf{Estimated}\\$(\hat{\phi},\hat{v})$} & \shortstack{\textbf{Deviation}\\$(\Delta\phi,\Delta v)$} & \shortstack{\textbf{Rel. Peak}\\\textbf{Power}} \\
\hline
T1 & $(0^{\circ},5.00~\mathrm{m/s})$ & $(0^{\circ},5.09~\mathrm{m/s})$ & $(0^{\circ},+0.09~\mathrm{m/s})$ & $0.00$~dB \\ \hline
T2 & $(10^{\circ},10.00~\mathrm{m/s})$ & $(10^{\circ},10.15~\mathrm{m/s})$ & $(0^{\circ},+0.15~\mathrm{m/s})$ & $-0.42$~dB \\
\hline\hline
\end{tabular}}
\end{table}

The measured angle--velocity map is shown in Fig.~\ref{fig:angle_velocity}. The relative peak powers reported in Table~\ref{tab:angle_velocity_results} are normalized to the strongest estimated target. Table~\ref{tab:angle_velocity_results} shows that both moving targets are estimated at their programmed angles. The estimated velocities are 5.09 and 10.15~m/s, with absolute deviations of 0.09 and 0.15~m/s. The small relative-power difference indicates that the two CE--PA--APM--PWG paths are sufficiently balanced in the present experiment. Together, the estimated angles and velocities demonstrate measurement of controlled Doppler and AoA states through the electronically steered DUT receive beam.

\section{Conclusion}
This paper presented a PWG-based OTA target-emulation method for compact multi-target and multi-angle sensing evaluation of phased-array devices. The method separates target-signal generation from spatial-wavefront synthesis. Delay, Doppler, and echo coefficients are controlled in the signal domain. Independent signal paths are mapped through a multi-input APM to different PWG-synthesized angular states over the same radiating aperture. This allows multiple target directions to be reproduced as local plane waves over the DUT aperture even when the test distance is shorter than the conventional far-field requirement. This extends compact PWG testing from plane-wave synthesis alone to multi-target and multi-angle sensing evaluation of phased-array devices.

The method was experimentally validated at 3~GHz using a real $1\times11$ phased-array DUT. The PWG--DUT separation was 1.6~m, substantially shorter than the 7.2~m Fraunhofer distance of the DUT. The synthesized incident fields exhibited amplitude PPEs of 1.0--1.1~dB and phase PPEs of $6.4^{\circ}$--$12.2^{\circ}$. Across the demonstrated angle--delay and angle--velocity cases, the recovered targets showed angular deviations within $1^{\circ}$, interpolated delay errors within 2.4~ns, and velocity errors within 0.15~m/s. These results demonstrate the feasibility of compact multi-target and multi-angle OTA sensing evaluation for large-aperture phased-array DUTs.

The present system is intended for emulating spatially close targets within a limited angular sector. The usable angular range is constrained by the finite PWG aperture, which reduce field-synthesis quality away from broadside. The number of simultaneously emulated target directions is limited by the number of independent input ports of the PWG-side APM.

The present experiments considered only fixed target AoAs. Dynamic-AoA emulation can in principle be realized by updating the PWG excitation vectors, but this capability has not yet been experimentally validated. Future work will evaluate dynamic-angle target emulation and investigate more accurate calibration and finer amplitude and phase control to improve emulation accuracy.

\bibliographystyle{IEEEtran}
\bibliography{refs}

\end{document}